\documentclass[]{interact}

\usepackage{epstopdf}
\usepackage{subfigure}

\usepackage{natbib}
\usepackage{hyperref}
\bibpunct[, ]{(}{)}{;}{a}{}{,}
\renewcommand\bibfont{\fontsize{10}{12}\selectfont}

\theoremstyle{plain}

\theoremstyle{definition}

\theoremstyle{remark}

\begin{document}


\title{Art Practice for Sustainability: A Cognitive-Affective-Systemic Framework}

\author{
\name{Ivan C. H. Liu\thanks{CONTACT Ivan Liu. Email: darknails@gmail.com Website: https://liu.lab.nycu.edu.tw LinkedIn: https://www.linkedin.com/in/ivanchliu/ ORCID: 0000-0002-9226-9765}}
\affil{Future Narratives Lab, Institute of Applied Arts, National Yang Ming Chiao Tung Univeristy, Hsinchu City, Taiwan (R. O. C.)}
}

\maketitle

\begin{abstract}
This paper proposes a \textit{Cognitive-Affective-Systemic} (CAS) framework that integrates cognition, emotion, and systemic understanding to cultivate sustainability awareness through art. Drawing from eco-aesthetics, affect theory, complexity science, and posthuman ethics, the framework defines artistic practice as both epistemic and performative---a way of knowing through making and feeling. Central to this is logomotion, an aesthetic mode where comprehension and emotion move together as a unified experience. Two artworks, \textit{SPill}, visualizing antimicrobial resistance through avalanche dynamics, and \textit{Echoes of the Land}, modeling anthropogenic seismicity, demonstrate how systemic modeling and sensory immersion transform complex science into embodied ecological understanding. The framework offers a methodological foundation for artists, theorists, and activists to translate awareness into engagement, advancing collective creativity toward sustainable futures.
\end{abstract}


\section{Introduction}

All events in the world are interconnected, forming an immense web of relations that links phenomena across space and time. The present state of the world emerges from innumerable past interactions, generating patterns of feedback, adaptation, and emergence that define complex systems. This systemic worldview central to complexity science and systems theory offers a powerful lens for understanding ecological, technological, and social phenomena as interdependent and continually evolving.

Within this intellectual lineage, cybernetics \citep{Wiener1950} provided the conceptual groundwork for exploring communication and control between humans, technologies, and environments. These feedback relations underpin today’s planetary condition, in which human and nonhuman actions are deeply entangled. The global emerging issues identified by the United Nations such as climate change, biodiversity loss, antimicrobial resistance, and social instability, illustrate how local perturbations can cascade through nonlinear systems, destabilizing ecological and social equilibria at global scales.

Scientific communication through data, graphs, and models makes these dynamics visible but seldom affective. Understanding alone does not ensure transformation. To confront planetary crises, knowledge must become embodied, affective, and participatory. Art occupies this crucial space. Through sensory experience, imagination, and affective resonance, artistic practice can cultivate ecological awareness that transcends data, activating empathy, curiosity, and care. As \citet{Duxbury2017} note, the arts play a vital role in sustainable development by ``raising awareness and catalysing action about sustainability and climate change.'' Artistic practice thus becomes not only representational but \textit{epistemic}: a way of knowing through making and feeling.

This paper draws from eco-aesthetics, affect theory, posthuman ethics, and complexity thinking to propose a three-dimensional framework, cognitive, affective, and systemic (CAS), for understanding and creating art that engages sustainability and complexity. The cognitive dimension treats art as a situated form of inquiry that translates systemic principles into experiential understanding. The affective dimension emphasizes art's capacity to evoke emotional and ethical resonance, transforming perception into embodied empathy. The systemic dimension positions art as both performative and analytical---enacting interdependence through aesthetic systems while probing the underlying mechanisms that govern dynamic interactions, enabling simulation, modelling, and experiential understanding of complex adaptive behaviors.

Together, these dimensions articulate a mode of practice in which art functions as an interface between cognition, emotion, and systemic awareness. By merging scientific reasoning with aesthetic experience, this framework advances a new methodology through which sustainability can be perceived, felt, and enacted. Within this framework, the paper introduces the notion of \textit{logomotion}---an aesthetic mode where cognitive insight and affective resonance move together as a unified process of understanding and feeling. Logomotion describes how the comprehension of systemic processes becomes itself an emotional experience, bridging intellect and empathy in the encounter with complexity.

To demonstrate this framework in practice, the paper presents two artworks as case studies: \textit{SPill}, a kinetic sound installation based on the sandpile model that visualizes the avalanche-like dynamics of antimicrobial resistance; and \textit{Echoes of the Land}, an interactive installation translating the spring-block earthquake model into a participatory multisensory experience of anthropogenic seismicity. Both works exemplify how artistic modelling and systemic aesthetics can make global complexity experientially tangible, linking cognitive insight with emotional and ethical engagement.

Beyond these examples, the CAS framework offers enduring value for artists, designers, theorists, educators, and activists seeking to address environmental and social challenges through creative means. For artists and designers, it provides a methodological foundation for developing works that integrate modelling, interactivity, and affective storytelling. For theorists and educators, it serves as a bridge between environmental humanities, complexity studies, and digital media. For activists and communicators, it offers strategies to transform scientific knowledge into participatory, emotionally resonant encounters.

This framework lays the groundwork for future creative and scholarly endeavors that merge systemic understanding with aesthetic expression. It positions art not as a supplement to sustainability discourse but as a vital mode of inquiry and action, where knowing and feeling, reason and emotion, converge in motion toward a more sustainable and empathetic planetary future.

\section{Theoretical foundations}
\subsection{Aesthetics and environmental humanities}

The growing urgency of climate change, biodiversity loss, and other global emerging issues has renewed interest in the role of art as a catalyst for ecological awareness and systemic understanding. Across the environmental humanities, art is increasingly theorized not merely as a representational tool but as a mode of knowing---a way to embody, experience, and reimagine complex planetary relations. 

Eco-aesthetic and phenomenological traditions ground this approach in sensory, affective, and relational experience. \citet{Berleant1992} and \citet{saito2017aesthetics} both emphasize immersion and participation, proposing that aesthetic experience collapses the distance between observer and environment, thereby nurturing ethical responsiveness. \citet{Morton2007} pushes this further, dismantling the romantic notion of a ``pure'' Nature and inviting us to confront the entangled coexistence of all beings. Within such frameworks, art becomes a practice of relational attunement, making perceptible the invisible feedbacks, dependencies, and contradictions that define life in the Anthropocene.


\subsection{Aesthetic cognition and experiential learning}

Philosophies of aesthetic cognition provide a parallel foundation for understanding how art can transform awareness into understanding. \citet{Dewey1934} situates art within the continuum of lived experience, turning perception into reflective knowing. \citet{Langer1953} extends this notion by describing art as a symbolic form---a medium capable of expressing patterns of feeling and thought beyond discursive language. In this view, art operates epistemically: it \textit{thinks} through making, complementing scientific abstraction with embodied, imaginative insight.

Pedagogically, this resonates with Freire's \citeyearpar{Freire1970} concept of \textit{conscientiza\c{c}\~{a}o}, in which creative and participatory practice fosters critical awareness and collective agency. Within sustainability discourse, it aligns with transformative learning \citep{Mezirow2000}, where aesthetic experience reorients perception and behavior toward ecological responsibility.

Complexity theory adds a systemic rationale to this cognitive-aesthetic lineage. \citet{Capra2014} describes life as networks of feedback and emergence beyond linear causality. Artistic practices—particularly generative, algorithmic, or agent-based ones—mirror these dynamics, offering intuitive access to systemic behaviors. Whitelaw's ``model worlds'' \citeyearpar{Whi2005_3} exemplify how data-driven or interactive artworks visualize flows and feedback loops underlying ecological and technological systems. The act of ``reading'' generative systems critically reveals their cultural and ethical underpinnings. Aesthetic theories of complexity \citep{Langton1990, Berlyne1968} further suggest that beauty arises at the \textit{edge of chaos}, where order meets surprise—a balance akin to ecological resilience. Through such synthesis, art becomes both representation and simulation: a cognitive bridge between systemic knowledge and perceptual experience.

\subsection{Affect theory and posthuman ethics}
If cognition allows us to \textit{understand} complexity, affect allows us to \textit{feel} it. The affective and ethical dimensions of sustainability-oriented art deepen this theoretical landscape. \citet{Massumi2002} describes affect as the pre-cognitive intensities, and sensation as the aesthetic manifestation of affect. Art and experience can make affect perceptible by activating sensation, bypassing rational interpretation. \citet{Ahmed2004} examines how emotional attachments organize collective life and how reorienting those attachments might create new possibilities for justice, solidarity, and transformation. Their theories together support the idea that art and experience can be drivers for social transformation, where the felt experience by audience creatives potential for further action.

As \citet{Bennett2010} argues, the liveliness of matter itself generates affective resonance, drawing humans into new relations with nonhuman forces. Such encounters can mobilize empathy, wonder, or discomfort—emotions that catalyze ethical awareness and potential behavioral change.

From an ethical perspective, Haraway's \citeyearpar{Haraway2016} notion of response-ability and Braidotti's \citeyearpar{Braidotti2013} posthuman ethics propose that responsibility emerges through relational entanglement rather than human exceptionalism. Haraway redefines ethics not as the moral obligation of an autonomous subject but as the capacity to respond within a web of co-becoming. To be ``response-able'' means to stay with the trouble---to inhabit the complexity of interdependence and vulnerability rather than seeking mastery or purity. Braidotti extends this position by arguing that the posthuman condition demands a shift from individualistic morality to a zoe-centered ethics, where life itself---human and nonhuman, organic and technological---is recognized as interconnected and vital. Both thinkers reject hierarchical divisions between subject and object, human and environment, proposing instead an ethics of immanent relations and mutual affectivity. Within this framework, art becomes a performative space where such ethics are enacted rather than merely represented. By giving agency to materials, algorithms, and ecological processes, artistic practice demonstrates how responsibility arises from participation and intra-action \citep{Barad2007}, not from external moral codes.

This relational ethics resonates strongly with sustainability's core principle of interdependence. When artworks invite audiences to experience feedback, transformation, or co-creation---whether through bio-art, environmental installations, or responsive media—they foreground the porous boundaries between human action and environmental change. The intertwining of affect and ethics thus positions art not simply as a medium of communication but as a relational practice that cultivates care, accountability, and imaginative co-existence within planetary systems.

\subsection{Cultural and narrative frameworks}

Complementing these philosophical and ethical foundations are cultural and narrative frameworks that situate art within broader social and communicative ecologies. Latour's Actor-Network Theory (ANT) and Facing Gaia \citeyearpar{Latour2017} conceptualize ecological awareness as the recognition of distributed networks of human and nonhuman actants. In Latour's model, agency is not confined to human intention but dispersed across hybrid assemblages---technologies, organisms, infrastructures, and institutions---that collectively produce the world. The ecological crisis, therefore, is not only environmental but ontological: a crisis of how humans conceptualize their place within a web of interdependent forces. Latour's call to ``compose the common world'' through new modes of representation and negotiation aligns with artistic practices that reveal these hidden networks of relation. Data visualization, interactive installation, and performative mapping can materialize the invisible flows of energy, information, capital, and emotion that bind ecological and social systems together.

Meanwhile, Heise \citeyearpar{Heise2008} reframes environmental consciousness through the concept of eco-cosmopolitanism, emphasizing the need to balance local attachment with global ecological awareness. For Heise, narratives---whether literary, cinematic, or artistic---mediate our sense of planetary belonging, enabling individuals to imagine themselves as participants in transnational ecological systems. Art, in this view, plays a vital role in cultivating a ``sense of planet'': an affective and cognitive recognition of shared planetary fate. By weaving together local data, personal stories, and global phenomena, eco-art and environmental media construct aesthetic bridges between the intimate and the planetary, the experiential and the systemic. This narrative dimension of sustainability art transforms awareness into empathy and global citizenship, encouraging audiences to see themselves as both place-bound and world-entangled beings.

\subsection{From narrative-centric art to social practice}

Finally, social practice theorists such as \citet{Kester2004}, \citet{Bishop2012}, and \citet{Lacy1995} expand this discourse by examining how participatory and dialogical forms of art foster collective reflection, civic imagination, and situated ethics. Kester's concept of dialogical aesthetics emphasizes conversation, empathy, and mutual understanding as the core of artistic meaning-making, positioning art as a social process rather than an object. \citet{Bishop2012}, while critical of overly utopian claims in participatory art, nonetheless acknowledges its political significance in reconfiguring relations between artist, audience, and institution, raising questions about authorship, agency, and power. Lacy's notion of new genre public art foregrounds collaborative, site-specific practices that merge aesthetic intention with social activism, often addressing gender, environmental justice, and community resilience.

Together, these frameworks highlight the civic and pedagogical potentials of art: to transform audiences from passive spectators into active co-creators of meaning and to create spaces for negotiation, empathy, and dialogue about shared ecological futures. Within sustainability discourse, such participatory practices extend the ethical and affective dimensions of art into the social sphere, where collective imagination becomes a precondition for systemic change.

Integrating these diverse perspectives, contemporary art operates as a cognitive-affective-systemic interface through which sustainability becomes perceptible, felt, and enacted. It does not merely depict ecological crises; it \textit{enacts} their dynamics. From immersive environmental installations and bio-art to algorithmic and participatory projects, such practices do not merely communicate ecological crises---they instantiate their dynamics. By visualizing data, enacting feedback, and invoking empathy, they make the complexity of the Earth system experientially tangible. In this sense, art's theoretical basis in eco-aesthetics, experiential learning, complexity, and posthuman ethics positions it as a vital mediator between scientific knowledge and lived awareness, offering not only reflection but the possibility of transformation

\section{The framework in a nutshell}
\subsection{The three dimensions}

\begin{figure}
\centering
\includegraphics[width=\textwidth]{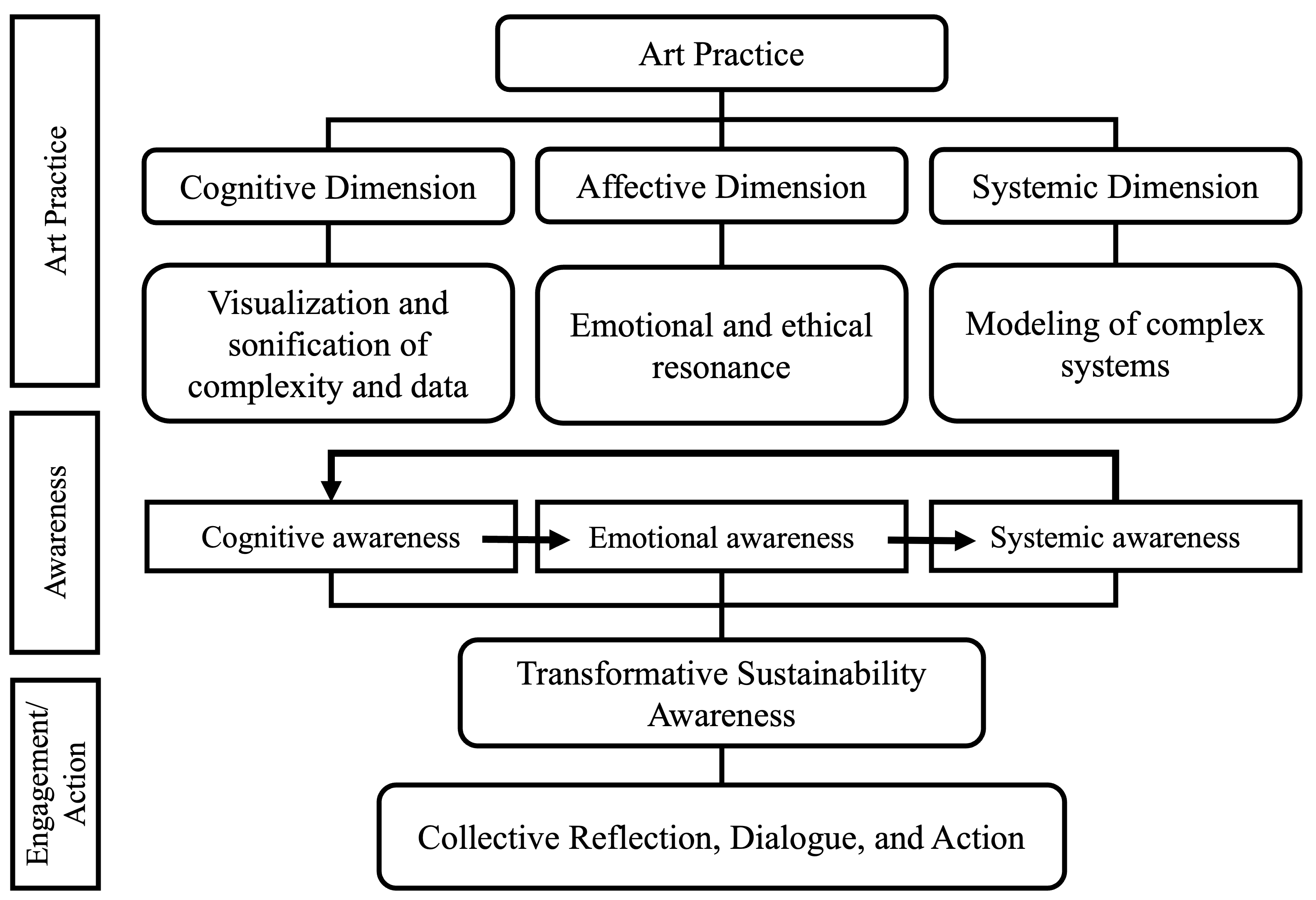}
\caption{\label{fig:conceptual} Conceptual diagram illustrating the theoretical framework linking art, awareness, and engagement/action. The diagram shows how art functions as an interface bridging cognition, emotion, and systemic understanding within the Cognitive–Affective–Systemic (CAS) model.\\
Alt Text: A conceptual diagram showing the relationship between art, awareness, and engagement/action. Art sits at the top as an interface bridging cognition, emotion, and systemic understanding. Three interlinked arrows form a cyclic loop between cognitive awareness, emotional awareness, and systemic awareness, while a vertical flow connects art practice to collective reflection and action.}

\end{figure}

Synthesizing these traditions, the proposed Cognitive-Affective-Systemic (CAS) Framework conceptualizes art as an interface connecting perception, emotion, and systemic understanding. As illustrated in Figure \ref{fig:conceptual}, it functions both epistemologically, as a way of knowing, and methodologically, as a guide for creative practice, that cultivates transformative awareness. It highlights both a cyclical relation among cognitive, emotional, and systemic awareness and a progressive movement from artistic experience toward social engagement and collective action.

At the top, art practice encompasses aesthetic, experiential, technological, and social aspects. It branches into three interrelated dimensions. \textit{Cognitive} dimension engages the intellect through visualization and sonification of complexity and data, drawing from disciplines such as, but not limited to, eco-aesthetics, complexity art, generative art, and experimental media art. It is the dimension related to aesthetics experience and the captivation of audience through experimenting artistic form, medium and representation. \textit{Affective} dimension evokes emotional and ethical resonance, informed by affect theory and posthumanism, emphasizing the embodied and ethical impact of art. This dimension connects the artistic experience to embodied human emotion. \textit{Systemic} dimension involves modelling of complex systems, grounded in complexity science, to reveal interdependencies within natural, ecological, and social systems. Concepts such as, but not limited to, cellular automata, self-organized criticality, and non-linear dynamics offer mathematical understanding of the underlying systemic mechanism that can be utilized as tools to translate scenarios into an artistic representation. This connects the scientific knowledge about the real-world complex issue to the generation of aesthetic experience.

These dimensions collectively generate three layers of awareness with a cyclic progression, from cognitive awareness, emotional awareness to systemic awareness and back to cognitive awareness, each building upon the previous. They encompass the mind, body, and ecology, forming a creative framework that is inherently more-than-human. This cyclical dynamic reimagines awareness not as a fixed state but as an evolving, relational process of co-becoming.

Vertically, the framework traces a trajectory from artistic creation, to awareness formation, and ultimately to collective reflection, dialogue, and action. At its foundation, collective reflection, dialogue, and action---informed by social practice, critical pedagogy, and eco-ethics---serve as pathways for translating artistic insight into participatory and transformative responses to sustainability challenges.

In practice, creative process starts with systemic research that informs modelling in systemic dimension. The artist then enters the cognitive dimension with the modelling tools---formalisms, program code, or devices---constructing an artistic representation through experimentation of media. Through audience perception of the representation, a kind of affective reaction ``brews'' in the affective dimension, merging the three dimensions giving rise to \textit{logomotion}, a mode in which cognition and emotion move together as a unified aesthetic process.

\subsection{Logomotion, embodiment and participatory aesthetics}
Derived from Latin word \textit{logos} (reason) and emotion (feeling), \textit{logomotion} proposes an aesthetic mode in which cognitive understanding and affective resonance move together as a dynamic unity. Within environmental and complexity-based art, logomotion describes how the comprehension of systemic processes---ecological feedback, emergence, or interdependence---becomes itself an embodied and emotional experience. Rather than separating intellect from feeling, it conceives awareness as a motion between the two: knowing as sensing, sensing as knowing. This dynamic synthesis echoes Dewey's view of art as experience and Massumi's notion of affective thinking, yet extends them toward systemic aesthetics. In this sense, logomotion redefines the sublime for the age of complexity: not as transcendence beyond understanding, but as wonder through understanding---an aesthetic cognition in which insight into the world’s unfolding order evokes empathy, care, and participatory awareness of our entanglement within it. Logomotion is an interdependent mixture of understanding and emotional resonance. No one comes before the other, and no one exits without the other.

Empirical evidence from neuroaesthetics supports this inseparability of comprehension and affect. In a landmark fMRI study, \citet{Zeki2014} demonstrated that the experience of mathematical beauty---elicited when mathematicians contemplated equations they personally regarded as elegant---activated the same region of the emotional brain, field A1 of the medial orbitofrontal cortex (mOFC), as the experience of musical or visual beauty. Crucially, the activity of this region correlated parametrically with the declared intensity of beauty even when the variable of understanding was statistically controlled. The findings reveal that intellectual reasoning and aesthetic pleasure converge within a shared neural substrate: comprehension itself can evoke the emotional signature of beauty. Zeki and colleagues thus provided physiological evidence for what philosophers since Plato have intuited---that understanding can be beautiful, and that beauty in turn signifies an alignment between cognition and affect.
 
Neuroscientific findings also reveal that this unity extends beyond cognition and emotion to embodiment. \citet{Kawabata2004} showed that when individuals viewed paintings they judged as beautiful, activation occurred not only in the mOFC but also within the motor cortex. Beauty and ugliness were associated with opposite modulations in this shared sensorimotor–emotional network: the mOFC responded positively to beauty, while the motor cortex showed greater activity for aversion. This suggests that aesthetic experience is inherently embodied, coupling perception, evaluation, and motor readiness within a single system. To perceive beauty is neurologically to move toward it; to perceive ugliness is to withdraw. Aesthetic judgment thus mobilizes bodily circuits of approach and avoidance, confirming that pleasure and aversion are enacted as much as they are perceived.

Together, these studies indicate that aesthetic experience arises from a neural loop linking reasoning, emotion, and movement. The mOFC integrates cognitive appraisal with emotional valuation, while the motor cortex grounds this synthesis in bodily readiness. This circuit parallels the systemic feedback dynamics central to the CAS framework: cognition triggers affective response, which in turn modulates perception and prepares action. The pleasure of insight arises from the recursive coupling of comprehension and embodied resonance---a form of knowing that moves through and with the body.

In the following sections, two artworks created within this framework, \textit{SPill} and \textit{Echoes of the Land}, illustrate how cognitive, affective, and systemic dynamics can converge in practice to foster sustainable awareness and engagement. 

\section{Case studies}
\subsection{\textit{SPill}: The avalanche of antimicrobial resistance}

Antimicrobial resistance (AMR) has emerged as one of the leading global public health threats in the 21st century \citep{UNEP2017}. It occurs when bacteria develop resistance to the drug diminishing its effectiveness. A recent global analysis published in The Lancet \citep{Murray2024AMR} estimated that in 2021, 1.14 million deaths were directly attributable to bacterial AMR, with an additional 4.71 million deaths associated with resistant infections. The study forecasts that by 2050, AMR could cause 1.91 million attributable deaths and 8.22 million associated deaths annually if current trends continue, underscoring its growing threat as one of the leading global health challenges of the century.

The mechanism of antimicrobials involves inhibiting or interfering with bacterial cell wall synthesis or protein synthesis, essential for bacterial survival. However, resistance to antimicrobials arises in bacteria through natural selection and mutation. Over time, certain bacteria can acquire genes that encode enzymes capable of breaking down antimicrobial agents before they can be effectively eliminated. This evolutionary process can be understood through the lens of complex systems theory: local genetic variations accumulate incrementally until the system reaches a critical threshold, after which resistance can spread rapidly through horizontal gene transfer and ecological interactions. In this sense, AMR exhibits characteristics analogous to an avalanche, where small, localized changes trigger large-scale and often irreversible cascades of resistance across microbial populations and ecosystems. Such nonlinear dynamics highlight the fragility of microbial equilibria under sustained selective pressure from antibiotic overuse.

Antibiotics can enter the environment through various pathways, including animal waste, human waste, and manufacturing waste. A notable contributor is the unwarranted use of antibiotics, wherein patients may insist on these medications from healthcare providers or acquire them over the counter without a comprehensive understanding of the implications or necessity. Additionally, improper disposal practices, such as flushing antibiotics down the toilet, further exacerbate the issue. Addressing pharmaceutical waste typically involves incineration, yet in many countries, the incineration temperature for regular household waste falls significantly below the necessary temperature ($1100^{\circ} C$) required for the effective breakdown of pharmaceutical wastes.

\begin{figure}
\centering
\subfigure[\label{fig:spill}]{%
\resizebox*{0.525\textwidth}{!}{\includegraphics{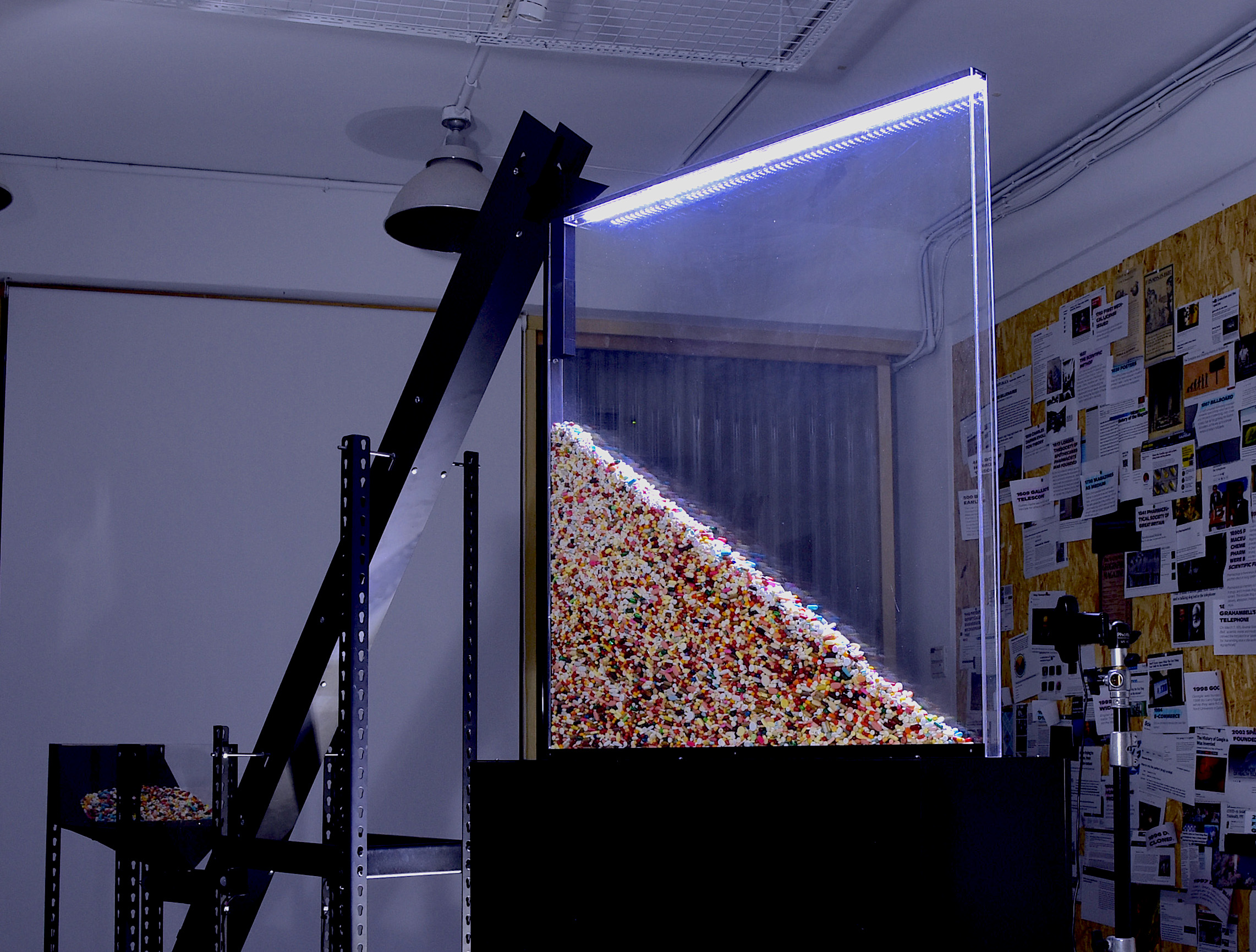}}}\hspace{5pt}
\subfigure[\label{fig:pills}]{%
\resizebox*{0.4\textwidth}{!}{\includegraphics{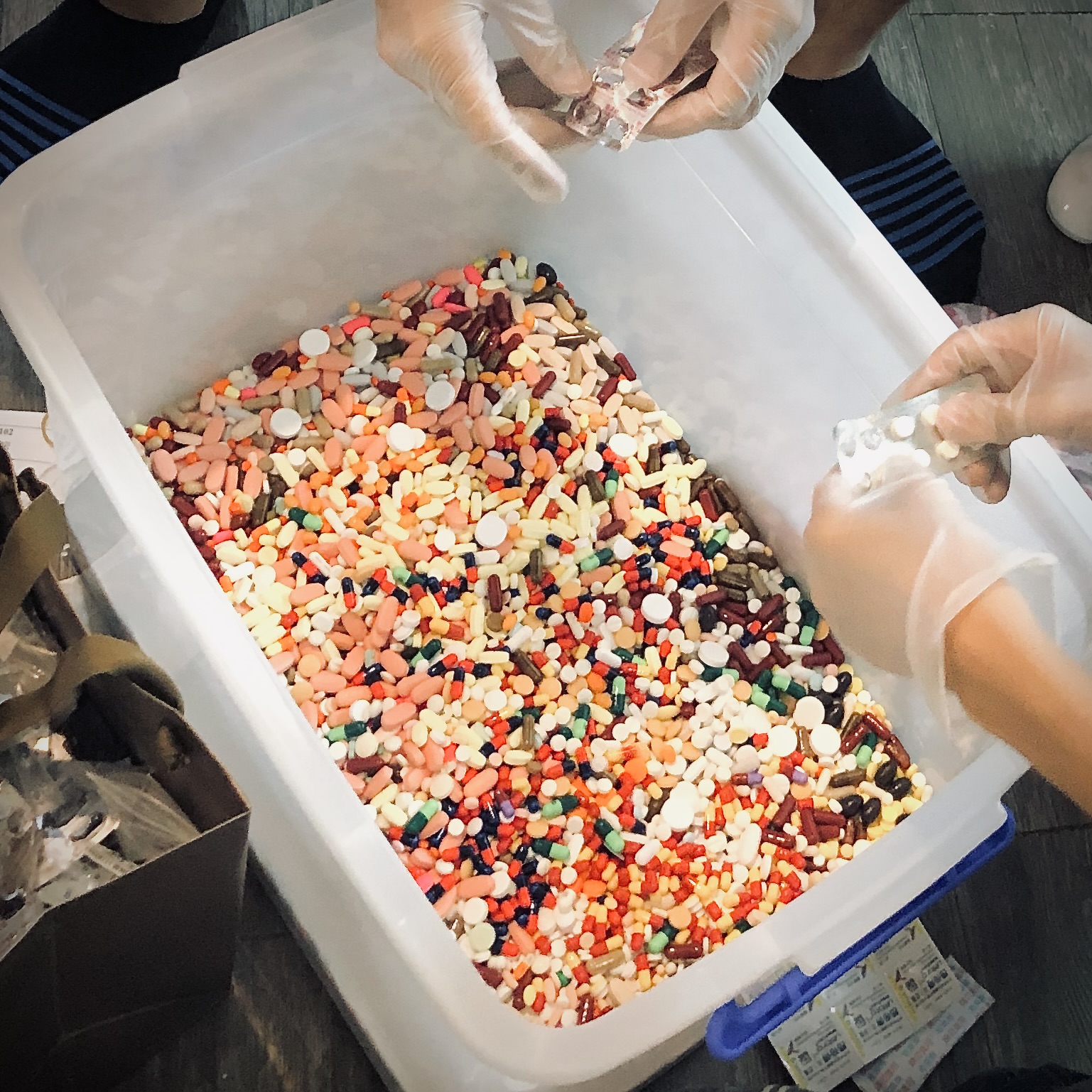}}}
\caption{(a) The kinetic sound installation \textit{SPill}, showing the main container filled with recycled pills forming a slope and the mechanical system that lifts and releases them to trigger avalanche events.  (b) Recycled pills collected and unpackaged for the installation.\\
Alt Text:
(a) A kinetic installation featuring a transparent container filled with colorful recycled pills forming a slope. A mechanical lifting system continuously moves the pills upward, creating periodic avalanche-like cascades.
(b) Close-up view of unpackaged recycled pills of varied colors and sizes arranged for use in the installation, highlighting pharmaceutical waste as artistic material.} 
\end{figure}

In order to raise the awareness about the AMR issue, we employed the physical mechanism behind the avalanche effect---the sandpile model---to construct the kinetic sound installation \textit{SPill}. The sandpile model was proposed in 1987 by physicists \citet{Bak1987} to illustrate the concept of self-organized criticality---how complex, scale-free behavior can emerge spontaneously in nature. In the model, grains of sand are dropped one by one onto a grid. When a site exceeds a critical threshold, it topples, distributing grains to its neighbors. This can trigger a cascade of further toppling. Over time, the system naturally evolves toward a critical state where avalanches of all sizes occur, following a power-law distribution without external fine-tuning.

Instead of sand, a quasi-one-dimensional avalanche machine was created using recycled pharmaceutical pills, as shown in Figure \ref{fig:spill}. A conveyor belt transports the pills to the top of installation and drop them into a thin large slab of transparent container until a critical slope is formed. This slope represents the equilibrium of the system. As more particles are released from the top of the pile, they are initially prevented from rolling down slope by local structures. They accumulate while more particles are released, until the load exceeds a threshold, and they roll downhill. Each time they roll downhill, the particles are dispersed and then stopped. However, if the process is repeated over time, the local landslide eventually turns into an avalanche.

To captivate audience in the cognitive dimension, we developed a real-time sonification technology that translate the micro-movement of the toppling pills into synthesized sound using a motion capture camera. The live video was fed to computer and processed, generating trembling sound effect using concatenative granular synthesis \citep{schwarz2006real} with the sound of iceberg breaking as audio input. This technique yields a highly naturalistic and impactful sonic experience that resembles a natural avalanche. The uncanniness between the scene of toppling pills and the sound of avalanche together creates a tension that captivates audience on a deep emotional level, achieving a logomotive state.

The title \textit{SPill} encapsulates multiple layers of meaning: it refers to the pill itself as an object; the literal act of spilling, evoking the cascading motion of pills toppling through the installation; the metaphorical overflow of excessive consumption; and the underlying sense of escalating, uncontrollable urgency.

The installation contains 10 kg of recycled waste pills (shown in Figure \ref{fig:pills}) donated by the public, whom we reached through digital social flatforms such as Facebook and Instagram. Beyond the description of our project, the social media posts disseminated pertinent information on safe disposal of pharmaceutical waste. This participatory approach transformed the creation process into a form of public engagement. The curated practical guidance and information motivated viewers to share the posts more willingly than usual. Notably, we observed approximately four-fold substantial increase in the number of shares and likes, demonstrating how intervention of the creation process can amplify public engagement and dissemination of sustainable knowledge.

The cascading motion captivates the viewers with a sense of simultaneous fragility and inevitability of the cascading motion. The visual aspect of the artwork also draws their attention---the vibrant colours, the avalanche, and the incredibly realistic avalanche sound generated by the movement of pills. Such curiosity and desire to participate motivated audiences to contribute their unwanted medicines to the installation. All the pills used in the work were ultimately recycled through pharmacies following proper disposal procedures.

By aligning systemic modelling with sensory experience, \textit{SPill} makes visible and audible the entanglement between individual behavior and planetary health. It transforms an abstract scientific concept into embodied ecological awareness, demonstrating how the cognitive, affective, and systemic dimensions of the framework converge in practice.

\subsection{\textit{Echoes of the Land}: The embodied perception of anthropogenic seismicity}

Anthropogenic seismicity, the triggering of earthquakes by human activity, illustrates another form of complex systemic feedback between society and environment. Activities such as wastewater injection, hydroelectric dam construction, and hydraulic fracturing have been documented to induce earthquakes across the globe \citep{Foulger2018}. Induced seismicity resulting from these activities has the potential to cause significant destruction to communities in the affected regions, posing a threat to the safety of human and non-human inhabitants. These events expose the fragile balance between technological expansion and geological stability, reminding us that the Earth is neither inert nor immune to human ambition. It is imperative to raise awareness about these issues among the public to foster a greater understanding and ensure the establishment of sustainable environments for future habitation.

\begin{figure}
\centering
\includegraphics[width=\textwidth]{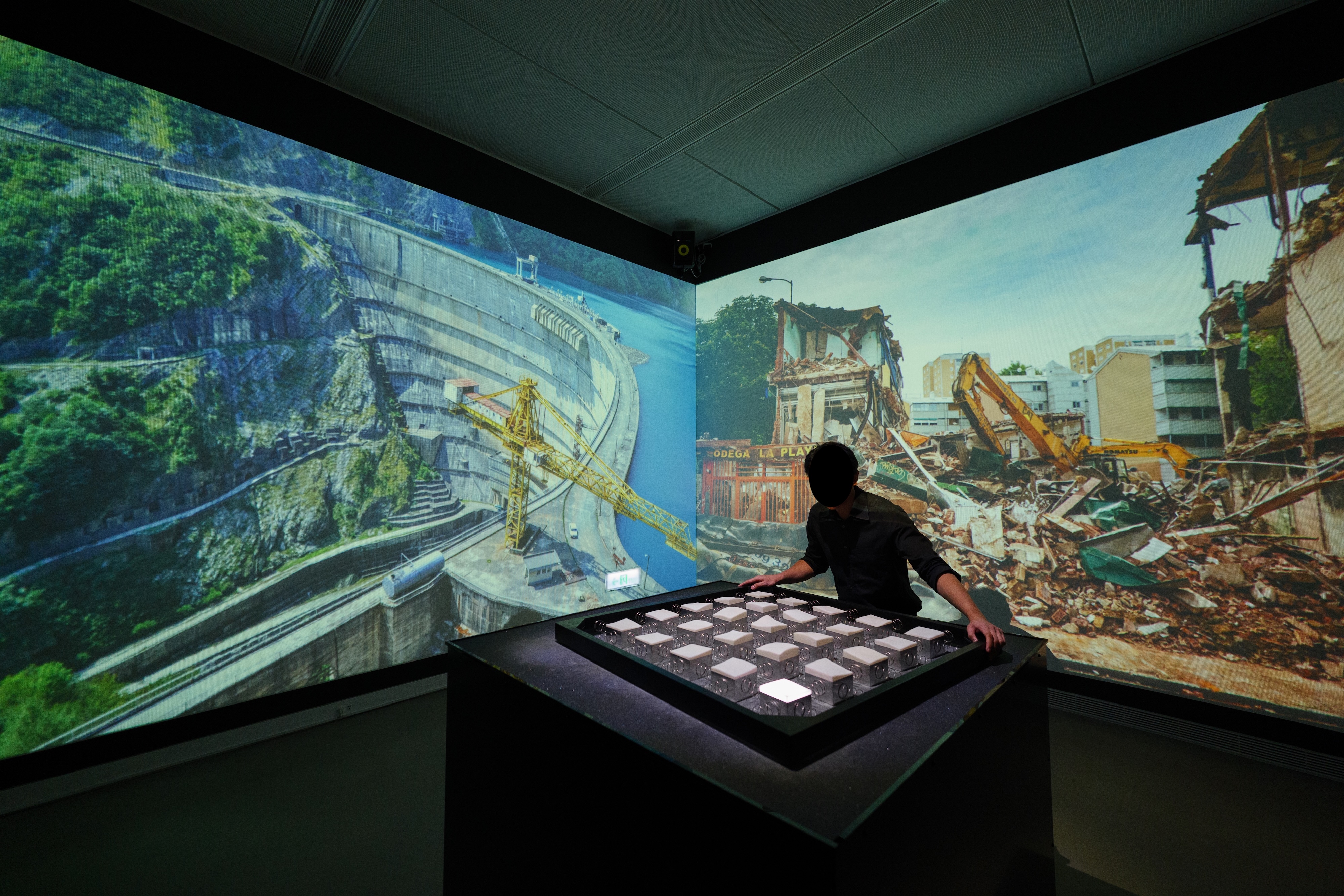}
\caption{\label{fig:EOL}%
The interactive multisensory installation \textit{Echoes of the Land}, showing the spring-block ensemble and immersive projection environment that translate human interaction into visual and sonic feedback.\\
Alt Text: A large interactive installation combining a grid of spring-blocks on a movable frame with immersive projection and sound. Participants push or move the frame, triggering lights and seismic-like sounds. Projected images of human infrastructure and environmental destruction appear on surrounding walls, linking human action to seismic consequences.
}
\end{figure}

The interactive installation \textit{Echoes of the Land} transforms this phenomenon into a multisensory environment that merges scientific modelling with participatory experience (Figure \ref{fig:EOL}). Its design is grounded in the spring-block model of earthquake recurrence \citep{Burridge1967,Brown1991}, which simulates tectonic stick-slip dynamics through a grid of masses connected by springs on a frictional surface. As stress accumulates gradually, it is released in sudden cascades once a critical threshold is surpassed—an archetype of self-organized criticality in geophysical systems.

\begin{figure}
\centering
\subfigure[\label{fig:BK}]{%
\resizebox*{0.42\textwidth}{!}{\includegraphics{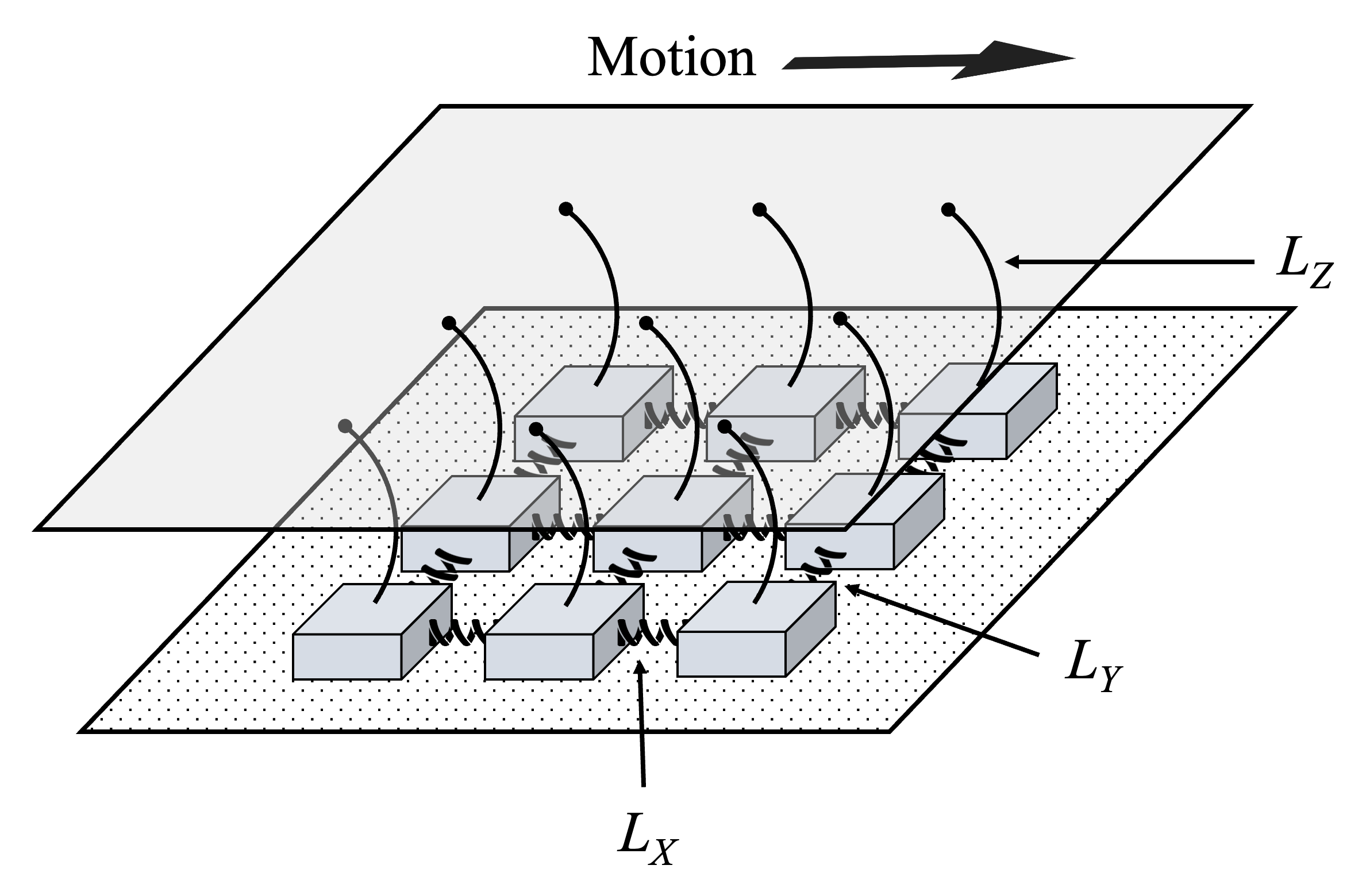}}}\hspace{5pt}
\subfigure[\label{fig:frame}]{%
\resizebox*{0.52\textwidth}{!}{\includegraphics{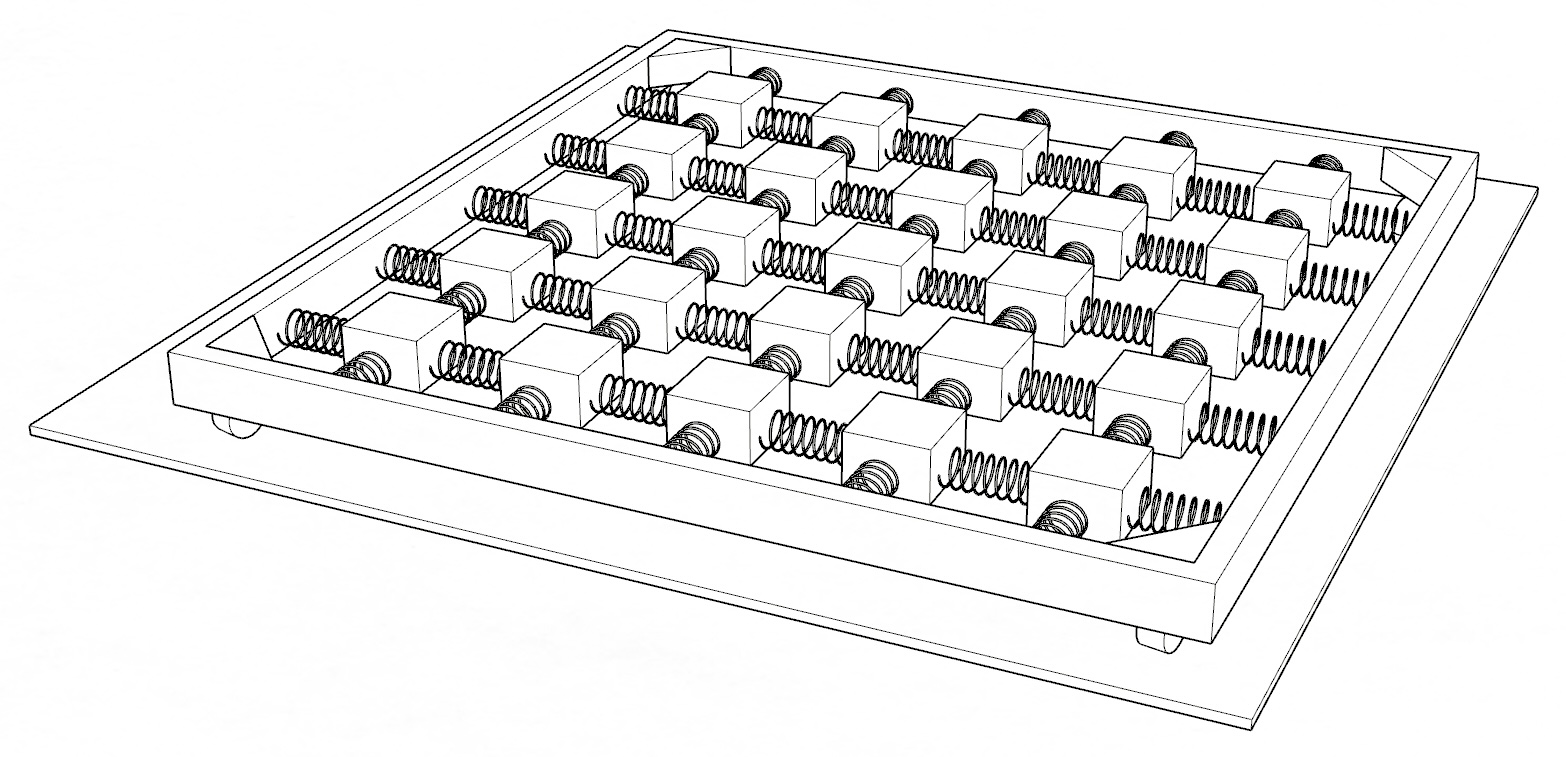}}}
\caption{(a) The spring-block model of earthquake recurrence proposed by \citet{Brown1991}. (b) Adapted $5\times 5$ spring-block system used in the installation \textit{Echoes of the Land}, enclosed within a movable frame for human interaction.\\
Alt Text:
(a) A schematic of the original spring-block model showing an array of interconnected blocks and springs simulating tectonic stress and slip.
(b) A physical $5\times 5$ grid of small square blocks joined by springs within a movable square frame, representing a tangible adaptation of the earthquake model for audience interaction
} 
\end{figure}

As shown in Figure \ref{fig:BK}, the model comprises 2D arrays of blocks interconnected by springs on a frictional surface. Each block is again connected to a suspended plate at the top through a spring, facilitating horizontal movement. Essentially, the blocks act as tectonic plates and the springs as the mediators for stress in-between. The earthquake recurrence is initiated by moving the top plate in a horizontal direction. The springs, $L_Z$'s, pull the blocks along with the plate. However, owing to friction between the blocks and the bottom surface, the majority remains stationary until the pulling force reaches a critical threshold, causing one block to slip. The slippage subsequently triggers a cascade of sequential movements propagating throughout the network of blocks via the springs $L_X$'s and $L_Y$'s, corresponding to a large-scale seismic event.

\begin{figure}
\centering
\subfigure[\label{fig:terrain}]{%
\resizebox*{0.48\textwidth}{!}{\includegraphics{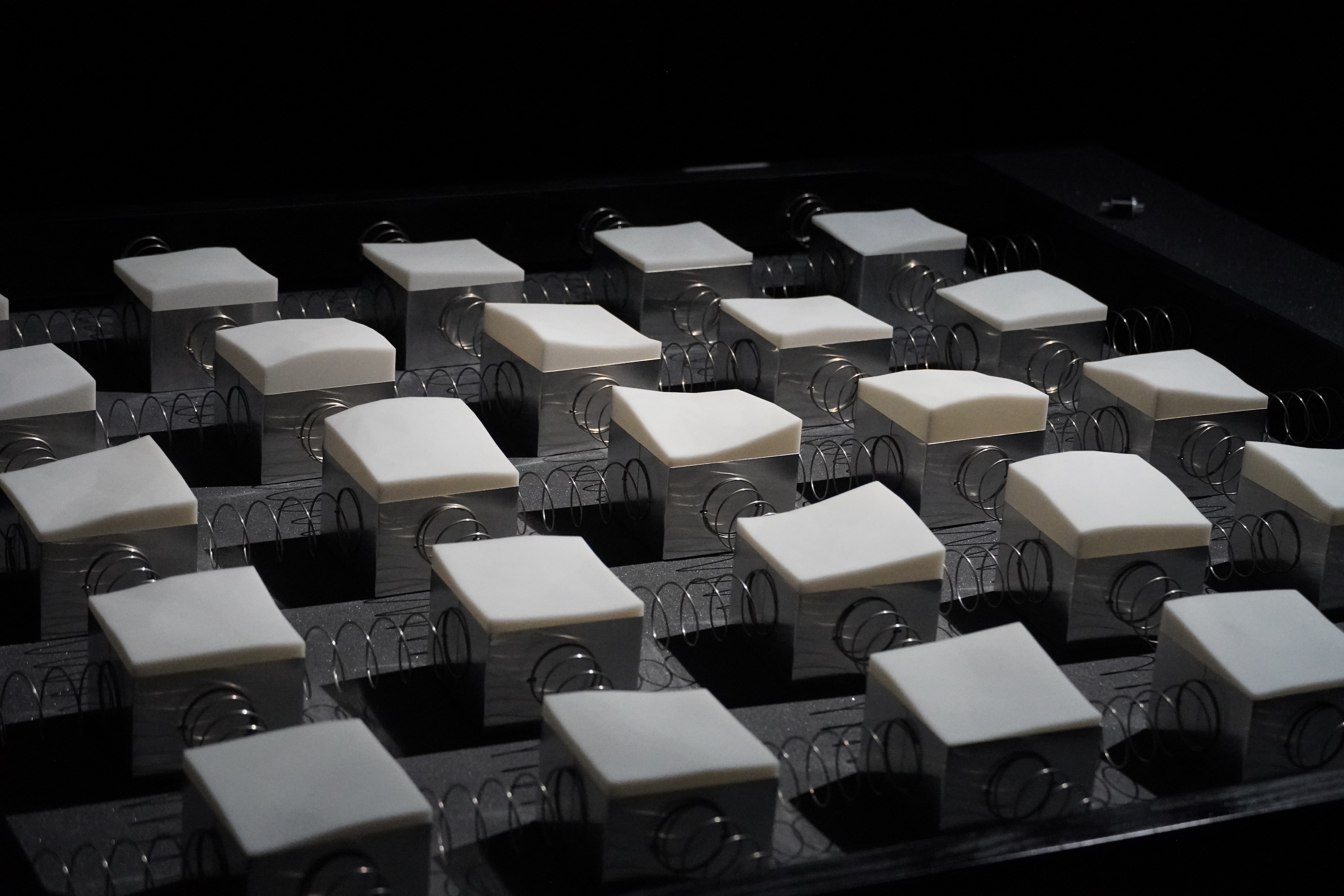}}}\hspace{5pt}
\subfigure[\label{fig:lights}]{%
\resizebox*{0.48\textwidth}{!}{\includegraphics{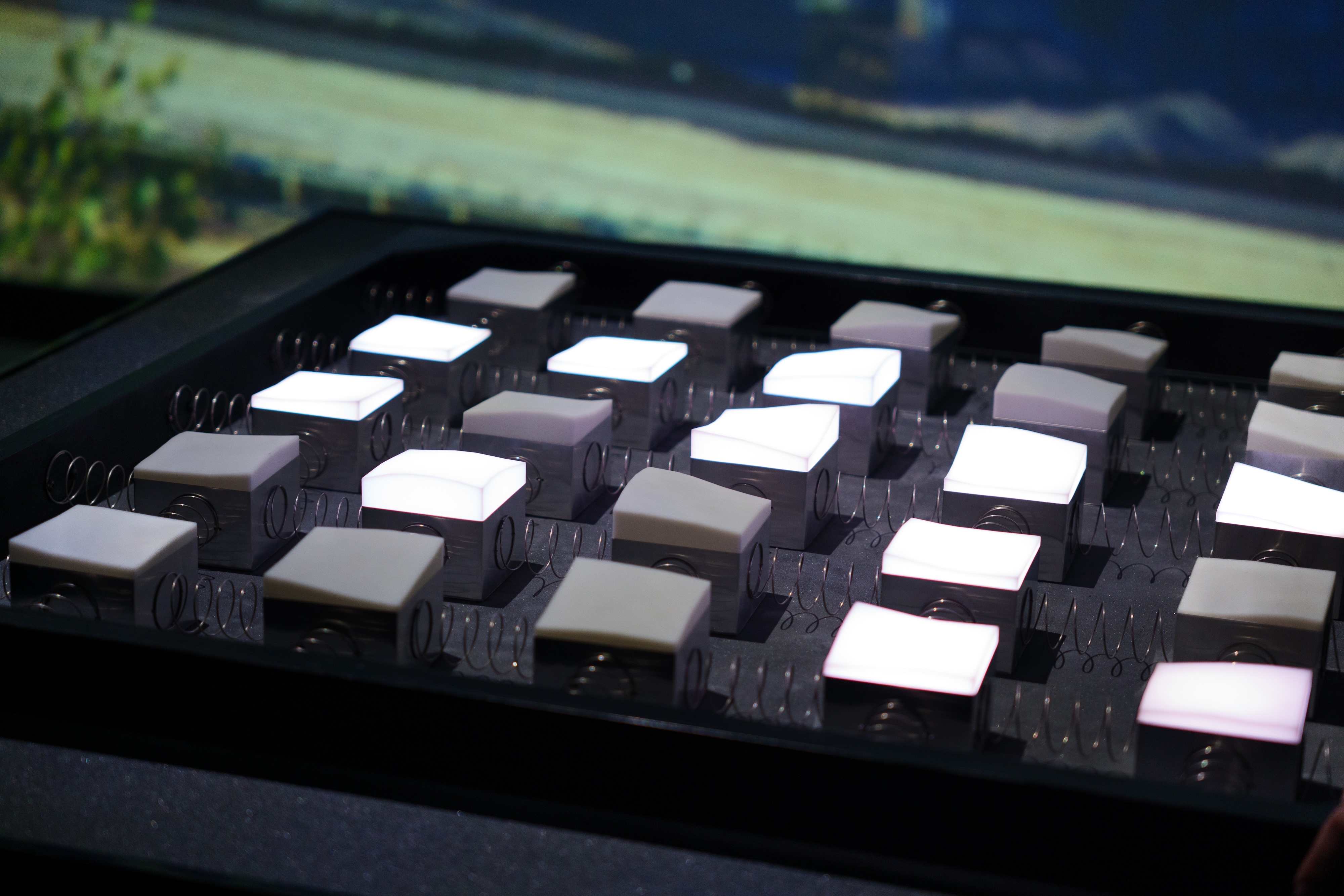}}}
\caption{\label{fig:terrainlight}(a) Close-up of the spring-block ensemble featuring an undulating terrain-like design. (b) Embedded accelerometers in each block trigger white flashes upon movement, visualizing seismic propagation.\\
Alt Text: (a) Detailed view of the spring-block structure resembling an undulating terrain, emphasizing the sculptural surface and interlinked motion system.
(b) Each block contains a small accelerometer; when movement occurs, bright white lights flash through semi-translucent covers, visualizing the propagation of simulated seismic waves..
} 
\end{figure}

In the artwork, this model is re-imagined as a $5\times 5$ spring-block ensemble enclosed within a movable square frame. Each block is connected by springs that allow limited horizontal motion, while the frame rests on caster wheels, permitting smooth, multidirectional movement. Audience can push, pull, or rotate the frame, generating stress within the network. When frictional limits are exceeded, the blocks slip, releasing motion that propagates through the array like seismic waves. Embedded accelerometers in each block detects these micro-movements, triggering flashes of white light through semi-translucent 3D-printed covers derived from digital terrain data Figure \ref{fig:terrainlight}. The visual effect suggests fragments of shifting landscape---an Earth responding to the pressure of human touch.

Similar to \textit{SPill}, a motion-capture camera mounted above the installation records each movement, while real-time concatenative granular synthesis converts the data into seismic-like sound, creating an aural field that vibrates between tremor and resonance. Surrounding the physical apparatus is an L-shaped projection space. On one wall, images of infrastructures known to induce earthquakes---dams, drilling platforms, skyscrapers---appear; on the other, their consequences---landslides, fractured roads, collapsed dwellings. The images shuffle rapidly with every detected movement and freeze when the system stabilizes, translating audience gestures into both cause and metaphor for anthropogenic disturbance.

As audience interact, they effectively generate their own earthquake data, enacting the feedback loop between human agency and geological response. The installation's tactile, visual, and sonic feedback transforms spectators into co-agents, allowing them to feel the thresholds and tensions that define the Earth system. This design intents to create an uncanny awareness of responsibility as the audience's small gestures triggered large-scale visual and sonic cascades, an affective embodiment of complexity itself. This uncanniness was intentionally created as a ``reflectaphor'' within the narrative structure. As described by \citet{Jennings1996}, it generates ``an irresolvable tension between similarity and differences.'' We utilized the reflectaphor to pique the audience's interest, curiosity, and emotions, with the intention of directing their attention towards the story conveyed by the works. Further detailed technical description and aesthetic analysis are published elsewhere \citep{Liu2025Anony}. 

\textit{Echoes of the Land} has been exhibited internationally at venues including Ars Electronica (2025), ISEA (2025), Bauhaus Museum (Weimar Kunstfest 2025), LEV Festival (2025), and IRCAM (2024), where it received enthusiastic critical reception. During these exhibitions, audiences not only experienced the multisensory impact of the installation but also learned about the scientific principles behind earthquakes through on-site demonstrations and explanatory materials. The work thus functioned as a form of science communication, translating seismological modelling into embodied and intuitive experience.

Across these presentations, the work was designed to prompt spontaneous dialogue among audiences, connecting the piece to analogous local phenomena in their own regions, underscoring the global resonance of induced seismicity.

Through its interactive design, \textit{Echoes of the Land} reconfigures the audience from observer to participant, transforming systemic knowledge into embodied awareness. Like \textit{SPill}, it demonstrates how the integration of scientific modelling, aesthetic experience, and collective interaction can render invisible planetary processes perceptible and emotionally compelling. Together, these two works exemplify the Cognitive-Affective-Systemic framework and the aesthetic mode of logomotion, where understanding and feeling converge in motion, turning environmental knowledge into empathy and participation.

\section{Conclusion}

In an era of ecological uncertainty and systemic entanglement, art stands as a vital form of inquiry that connects cognition, emotion, and system awareness. Building on the theoretical foundations of eco-aesthetics, affect theory, complexity science, posthuman ethics, and experiential learning, this paper proposed the Cognitive-Affective-Systemic (CAS) framework—a model that positions artistic practice as both a mode of inquiry and enactment. Through this framework, art is redefined not as representation but as a dynamic interface that models and performs the feedback relations shaping planetary life. It offers a methodology for knowing through making and feeling, where scientific reasoning, aesthetic experience, and ethical reflection converge to generate embodied understanding.


At the heart of this framework lies the new aesthetic mode of \textit{logomotion}, a synthesis of logos (reason) and emotion (feeling), which redefines aesthetic experience as a dynamic interplay between comprehension and affective resonance. Logomotion describes the continuous movement between understanding and emotion, in which cognition becomes sensorial and feeling becomes a mode of knowing. It encapsulates an aesthetic cognition where insight itself evokes empathy and ethical reflection, turning systemic awareness into lived experience.

Neuroscientific research supports the existence and significance of the logomotive state. \citet{Zeki2014} showed that experiencing mathematical beauty activates the medial orbitofrontal cortex (mOFC), linking comprehension directly to emotional reward. Similarly, \citet{Kawabata2004} found that perceiving artistic beauty engages both the mOFC and motor cortex, indicating that aesthetic experience is inherently embodied and participatory. Together, these studies suggest that beauty arises not only from perceptual or intellectual harmony but through active, embodied engagement where understanding, emotion, and action converge.

The artworks \textit{SPill} and \textit{Echoes of the Land} exemplify this approach. Created within the CAS framework, they employ systemic knowledge---drawing from avalanche dynamics and spring-block model---to construct naturalistic and intuitively comprehensible experiences of global phenomena. Their forms unfold through feedback, emergence, and self-organization, mirroring the behavior of the systems they represent. This naturalistic coherence makes their messages immediately legible: audiences can feel complexity as much as they can grasp it. By merging physical modelling with sensory immersion, these works transform abstract data into lived, participatory understanding. The integration of scientific mechanisms does not obscure but clarifies---the artworks become perceptually coherent, emotionally resonant, and intellectually engaging. Through this process, participants enter a \textit{logomotive} state in which feeling and understanding flow as one.

As audiences physically and emotionally engage with these works, sustainability awareness deepens the aesthetic experience: understanding one's entanglement with ecological systems heightens empathy, attentional focus, and pleasure. In this way, sustainability engagement and embodied action amplify the neural and experiential dimensions of beauty, turning comprehension into resonance and reflection into ethical delight.


The CAS framework provides a robust theoretical and methodological foundation for artists, designers, theorists, educators, and activists seeking to integrate systemic modelling, affective resonance, and engagement. It invites interdisciplinary collaboration across art, science, and the humanities, cultivating practices that transform sustainability from abstract discourse into embodied awareness and collective action. By uniting cognition, affect, and system through logomotion, this framework lays the groundwork for a new generation of creative inquiry, one that not only represents sustainability but enacts it, guiding artists, scholars, and activists alike toward a more responsive, relational, and sustainable planetary future.


\section*{Acknowledgements}
The author thanks the assistances of Alven Chen, Daniel Boubet, and Wei-Tzu Tseng for realizing \textit{SPill}; Chung-En Hao and Jing Xie for realizing \textit{Echoes of the Land}.

\section*{Disclosure statement}
The author reports no potential conflict of interest.

\section*{Funding}
This work was supported by the National Science and Technology Council (Taiwan) with grant no. MOST-111-2420-HA49-005 and NSTC 112-2420-H-A49-002.

\section*{Notes on contributor}
Ivan Liu is an artist and a researcher. He studied physics at Imperial College London, and obtained PhD from Max-Planck Institute for the Physics of Complex Systems in Germany. He was an independent artist for 10 years before becoming a full-time faculty member at the National Yang Ming Chiao Tung University in Taiwan, where he founded the Future Narratives Lab. His current works explore new ways to narrate contemporary issues with science-inspired artistic representations. His works have exhibited worldwide including Ars Electronica, ISEA, Bauhaus Museum Weimar and IRCAM.

\bibliographystyle{tfcad}
\bibliography{CASrefs}
\end{document}